\documentclass[aps,prl,twocolumn,showpacs,preprintnumbers,amsmath,amssymb,fleqn]{revtex4}
\usepackage{graphicx}
\usepackage{dcolumn}
\usepackage{bm}
\usepackage{subfigure}
\begin{document}

\preprint{COSM-06-02/LIPSS-06-015 v4}

\title{Production and Detection of Very Light Spin-zero Bosons\\at Optical Frequencies}

\author{A. V. Afanasev}
\author{O. K. Baker}
\email{oliver.baker@hamptonu.edu}
\homepage{http://cosm.hamptonu.edu/}
\author{K. W. McFarlane}
\affiliation
{Department of Physics, Hampton University, Hampton, Virginia 23668, U.S.A.}%
\author{G. H. Biallas}
\author{J. R. Boyce}
\author{M. D. Shinn}
\affiliation{Jefferson Lab, 12000 Jefferson Avenue, Newport News, VA 23606, U.S.A}

\date{\today}

\begin{abstract}
The PVLAS collaboration has observed rotation of the plane of polarization of light
passing through a magnetic field in vacuum and has proposed that the effect
is due to interaction of photons with very light spin-zero bosons. This would
represent new physics beyond the Standard Model,
and hence it is of high interest to test this hypothesis.
We describe a proposed test of the PVLAS result,
and ways of producing, detecting,
and studying such bosons with light in the optical frequency range.
Novel features include methods for measurements of boson mass,
interaction strengths, and decay- or oscillation-lengths
with techniques not available in the x-ray region.

\end{abstract}


\maketitle

%
\section{Introduction}
\label{sec:introduction}
The PVLAS collaboration's measurement\cite{Zavattini:2005tm},
of the rotation of the plane of polarization of light passing
through a magnetic field in a vacuum, is
many orders of magnitude larger than
expected from light-by-light scattering in
quantum electrodynamics~\cite{HeisenbergEuler:1936}.
They suggest that this effect is due to the
existence of a very light, neutral, spin-zero boson (LNB) of mass
1\,meV~$\lesssim m_b \lesssim 1.5 $\,meV,
coupling to two photons with a coupling constant $g_{b \gamma \gamma} = 1/M_b$, such that
$2 \times 10^5$~GeV~$\lesssim M_b  \lesssim 6 \times 10^5$\,GeV.
This boson would have properties similar to the
previously proposed axion~\cite{Weinberg:1978, Wilczek:1978},
including very weak coupling to other particles.
The ranges quoted for $m_b$ and  $M_b$
come from combining the PVLAS result with results from
previous experiments by the BFRT collaboration
\cite{Cameron:1993mr, Cameron:1992cb, Ruoso:1992nx}.

The existence of such a LNB would have many implications.
It would be a candidate for a component of dark matter.
It is not in the parameter range expected
for an axion~\cite{PDBook}, so it may indicate a new mass scale
in particle physics~\cite{Masso:2005}.  There may be more than one member
of the family,
with other unexpected properties \cite{Dienes:2000}.
Confirmation or refutation of this interpretation of the PVLAS result
is of high importance.

Since the effect has not been observed in astrophysical studies~\cite{PDBook},
a laboratory experiment is the most direct test.
Several groups are planning to carry out such
experiments (see, e.g., \cite{Rabadan:2005dm, Duvillaret:2005}).
There are advantages in using light in the visible and near-visible
frequency range: high-power coherent sources are available (such as the Jefferson
Lab FEL~\cite{JLab:FEL}),
as are a variety of sensitive detectors, and signals can be manipulated
with optical techniques.

\begin{figure}[h]
\includegraphics[keepaspectratio, width=3.375 in]{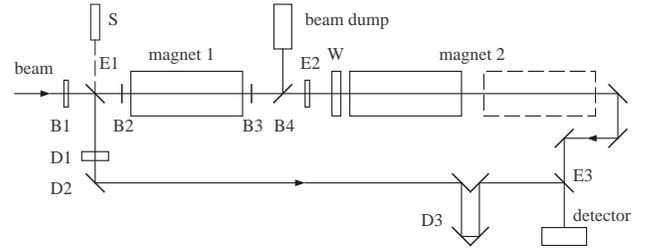}
\caption{Schematic layout of apparatus showing input laser beam,
LNB generation magnet 1, photon regeneration magnet 2, and
detector.  The magnetic fields are transverse to the beam direction.
The main beam elements are a polarization
rotator B1, a possible optical cavity (B2,B3),
a turning mirror B4 that also transmits a fraction of the beam
for alignment purposes, and a ``wall" W that prevents primary
laser light from reaching the regeneration magnet. Optical elements
E1-E2-E3 and E1-D1-D3-E3 form an interferometer.}
\label{fig:apparatus}
\end{figure}

Here we consider methods for production, detection, and study, of
LNBs in the specific mass and coupling constant
range of the PVLAS result.  Ideas behind the methods were put forward
in Refs. \cite{Sikivie:1981, VanBibber:1987} and some were employed
by the BFRT collaboration.  The approach uses production 
of LNBs by light in a static magnetic field (generation), followed by
conversion of LNBs to photons (regeneration), also in a static magnetic field.
The generation-regeneration (G-R) apparatus that we consider is shown
in Figure~\ref{fig:apparatus}.
The regeneration magnet can be moved along
the axis of the apparatus.

If the coupling of the LNB to ordinary matter is extremely weak,
the regenerated photons will have the same optical characteristics
as the original beam, except for a phase shift due to the mass
of the intermediate state, and hence should have predictable optical
proporties, such as being focusable to a very small area.
A low-noise imaging detector can be used, where the photons are
focused on one or a few pixels, the remaining pixels
giving an automatic background measurement.
An imaging detector can also observe beam properties other than rate.

The probability of a photon energy $\omega$ generating a LNB mass $m_b$
in a homogeneous magnetic field $B_1$ length $\ell_1$, followed by 
regeneration by the LNB of a photon in a field $B_2$ length $\ell_2$
 can be written, using notation and units
as in Refs.~\cite{Raffelt:1988,Ruoso:1992nx,Zavattini:2005tm}, as:
\begin{equation}
P_{GR} =
\left(\frac{B_1 \ell_1}{2 M_b}\right)^2\frac{sin^2(y_1)}{y_1^2}
\left(\frac{B_2\ell_2}{2 M_b}\right)^2 \frac{sin^2(y_2)}{y_2^2}
\label{eqn:G-R1}
\end{equation}
where $y_{1,2} = \Delta_b \ell_{1,2}/2$, and
when $\omega \gg m_b$, $|\Delta_b| = m_b^2/ 2 \omega$ in vacuum.

The interaction depends on the the photon's polarization,
the direction of the magnetic field, the parity of the LNB, and the parity
properties of the interaction.
For a pseudoscalar LNB, Eqn.~\ref{eqn:G-R1} applies when the
photon polarization is parallel to the field of the generation magnet,
the photon direction is perpendicular to the field,
and the interaction is parity conserving.  With similar provisos,
the polarization of the regenerated photons will be parallel to the
field of the regeneration magnet.  Rotation of the initial
polarization, or measurement of the polarization of regenerated
photons, relative to the direction of the relevant magnetic field, 
will reveal the combination of the
parity of the boson and the parity-conserving properties of the interaction.

The probability has zeros at $\Delta_b \ell_{1,2} / 2 = \pi, 2 \pi, ...$
and rapidly declines beyond the first zero, so that effective G-R
requires $\Delta_b \ell_{1,2} /2 < \pi$.
The PVLAS result 
sets a scale for constant-field G-R experiments:
the product $\ell_{1,2} \lambda$ should be $\lesssim 1$\,m$\cdot \mu$m,
or $\ell_{1,2}/\omega \lesssim 1$\,m/eV.
Magnets with periodic fields~\cite{Sikivie:1981} avoid this limitation
and rates can grow as the fourth power of the length of the apparatus.

\begin{figure}[h]
\includegraphics[keepaspectratio, width=3.375 in]{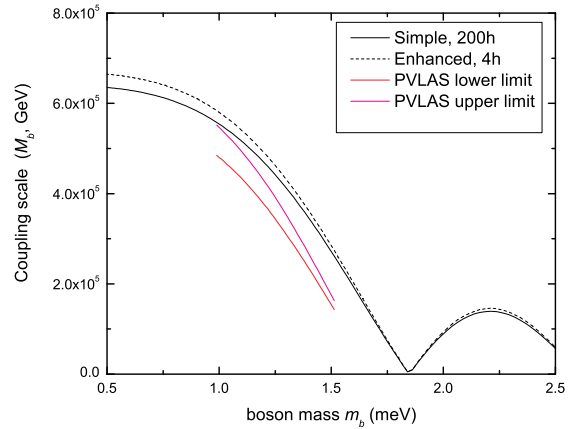}
\caption{Coupling scale performance of proposed apparatus, versus LNB mass,
compared with the PVLAS upper and lower 3-$\sigma$ limits.
The `Simple' curve gives the 5-$\sigma$ confirmation level for a 200-h run,
while the `Enhanced' curve is for a 4-h run using the signal enhancement
technique described here.}
\label{fig:lipss-reach}
\end{figure}

For a constant-field G-R arrangement, and
a rate $N$ of photons traveling in the beam direction in an
apparatus with overall efficiency $\eta$,
the number of regenerated photons observed at rate $r$ in time $t$
is 
$ rt = N t \eta P_{GR}$.
With a noise or background rate $n$, 
the significance reached is
\( S = r\sqrt{t/n} \)
where the variance of the background rate is taken as $n$.
A search experiment should achieve $S \geqslant 5$. 
As an example, two 1.0-m, 1.5-T magnets with 3\,kW of beam at 900\,nm and
a low-noise ($\sim 0.1$\,s$^{-1}$) detector,
would reach the $S \geqslant 5$
confirmation level in the PVLAS region in about 200 hours,
as shown in Figure~\ref{fig:lipss-reach}.

%
\section{\label{sec:PSE}Phase measurement and signal enhancement (PSE)}

\begin{figure}
\includegraphics[keepaspectratio, width=3.375 in]{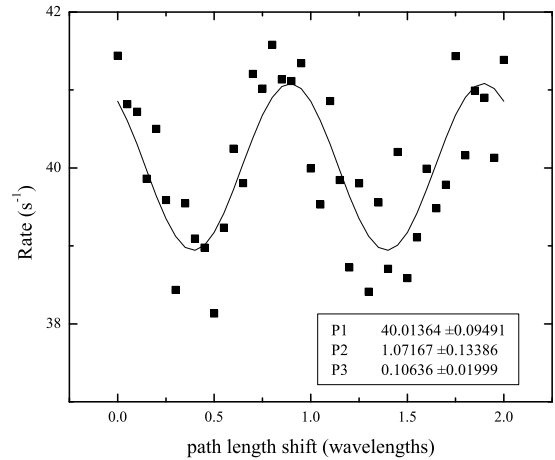}
\caption{Sample simulation of use of PSE method.  Points are simulated data, curve
is least-squares fit.}
\label{fig:pse-sim}
\end{figure}

The signal-to-noise ratio can be enhanced
using interference between regenerated photons and light from 
the incident beam~\cite{VanBibber:1987}.  Measurement of the phase
can also result in a mass measurement.
Figure~\ref{fig:apparatus}
includes a PSE device; in format it is a Mach-Zender interferometer.
This technique relies on the interaction of
the LNB with matter being small enough, and its lifetime
long enough, that the regenerated photons have
a phase relationship to the generating photons.  If
a signal were observed but interference was not,
that would indicate that the LNB has significant interactions.

Adding a reference beam of amplitude $A$
to a LNB amplitude $a$ with relative phase $\phi$ 
will result in a combined observed rate
$n + \eta (A^2  + 2aA\cos \phi  + a^2 )$.  The signal
rate without enhancement would be $r = \eta a^2$.
By varying the length of one arm, with the adjustable element D3,
the relative phase and amplitude can be measured.
For a simple estimate of the signal enhancement,
take the effective signal to be $2 \eta a A$, and the rate from
the reference beam to be  $\sim\! 3n$.
The significance obtained in time $t$ is now
$S'  = \sqrt {3 r t}$.
As noted in Ref. \cite{VanBibber:1987}, this result is noise-independent.
For a signal-to-noise ratio of $\sim\! 1:100$, the improvement in significance will
be $\sim\!20$.
Even in the absence of background or noise, a significance enhancement
of a factor of two can be obtained, equivalent to a reduction of time by
a factor of four.

A simulation of the process was made using $r = 0.01$\,s$^{-1}$
and $n = 10$\,s$^{-1}$, with a relative phase of 0.1 periods.
Runs consisted of 41 data points each corresponding to 100\,s of data.
The result of one such simulated run is plotted in
Figure~\ref{fig:pse-sim}.
An ensemble of runs gave a mean for the modulation amplitude
of $(1.13 \pm 0.15)$\,s$^{-1}$, compared with 1.1\,s$^{-1}$ expected.
The mean fitted pathlength shift was $(0.10 \pm 0.012)$ wavelengths.
(Both uncertainties are rms deviations of the individual run results from the
mean of all runs.)

The LNB mass can be measured by changing the spacing between
the generation and regeneration field centers and measure the resulting
phase shift. 
For example, the simulated results described above imply that a 1-m
movement of the regeneration magnet with 900-nm light
would give measurement of the mass
of a 1\,meV boson to $\sim\!0.02$\,meV.

%
\section{\label{sec:periodic}Periodic fields}

The use of periodic fields to detect and produce axions has been 
proposed~\cite{Sikivie:1981, VanBibber:1987}.
Periodic-field magnets can be extended as far as beam divergence allows,
overcoming the limitation $\ell \leqslant 2 \pi/\Delta_b$.
We propose novel ways of implementing
these ideas with magnets with adjustable periods, that
can be used to make precise measurements of LNB mass
and resolve components if there is more than one LNB.
For masses in the PVLAS region and photons near the visible range,
the required period is of the order of meters, making it
practical to consider periodic fields for a variety of
experiments (and for LNB production facilities).
In the x-ray region, the required period is of the
order of kilometers.

\begin{figure}
\subfigure[]{%
\includegraphics[keepaspectratio, width=1.45 in, height= 1.2 in]{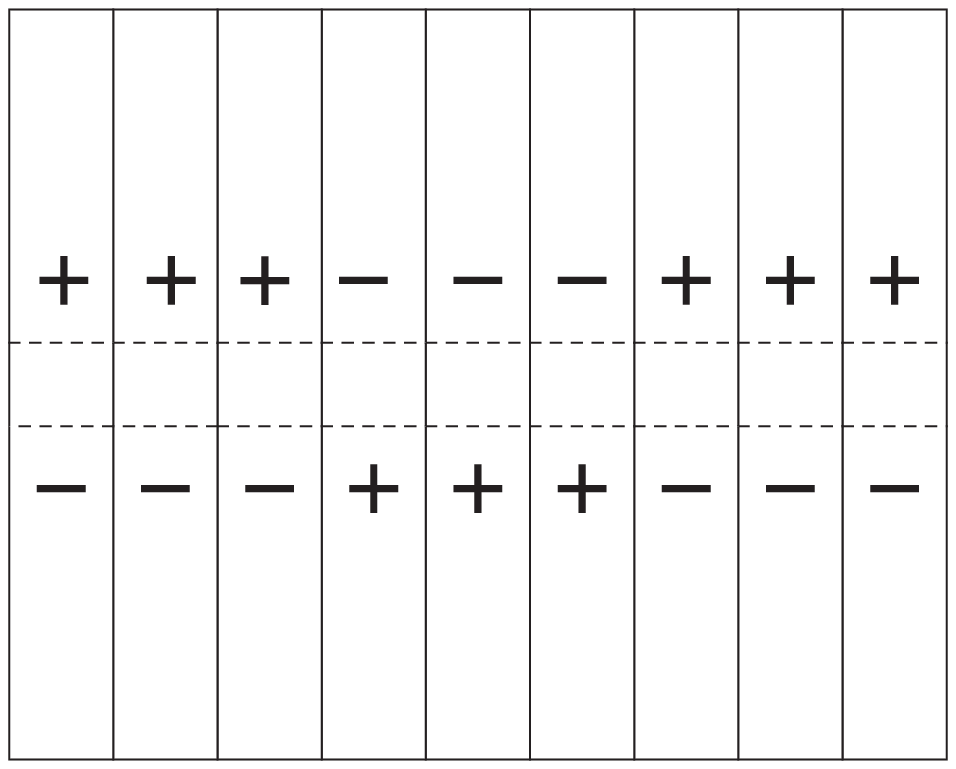}%
\label{fig:alternating}}
\subfigure[]{\includegraphics[keepaspectratio, width=1.2 in, height=1.2 in]{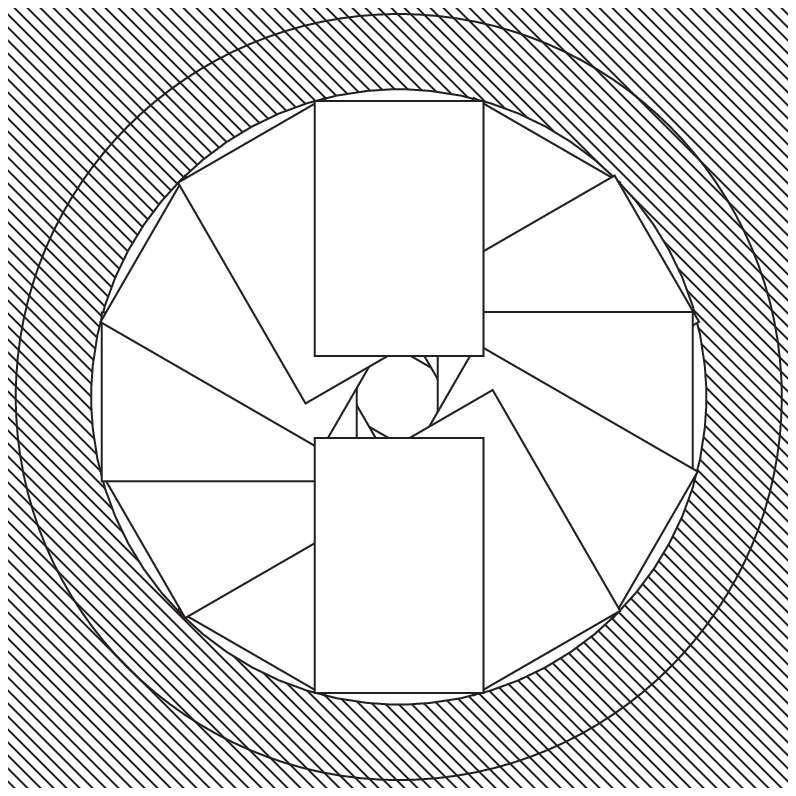}
\label{fig:helical}}\\%
\caption{\label{fig:periodic_magnets} Example of magnets to produce periodic fields:
(a) alternating magnetic fields produced by rotatable magnet segments,
(b) similar magnets used to form a helical field.}
\end{figure}

Figure~\ref{fig:periodic_magnets} shows
examples of how an adjustable periodic field could be produced.
The magnet in
Figure~\ref{fig:alternating} consists of rotatable elements that can
be arranged with alternating parallel fields, or at any
angle to form a helical field, as in Figure~\ref{fig:helical}.

We extend the alternating-field concept~\cite{VanBibber:1987}
to segmented magnets with periods incommensurate with the segment length. 
The probability of generation or regeneration in a magnet with alternating field
magnitude $B_m$ is:
\begin{equation}
P'_{G,R} = \left(\frac{B_m d}{2M_b}\right)^2 \frac{\sin^2{y_0}}{y_0^2}
\left|\sum_{k=1}^{n}\delta_k \exp{i(2k-1) y_0}\right|^2
\end{equation}
where $y_0 = \Delta _b d/2 = m_b^2 d/4 \omega$, $d$ is the length of each of
$n$ segments, and $\delta_k = \pm 1$ indicates the polarity of the $k$th segment.
The width of the resonant production is
$\delta m_b \approx 2 \omega / m_b \ell$,
with the maximum occurring at a period of 
$4\pi \omega /m_b^2 $.

\begin{figure}
\includegraphics[keepaspectratio, width = 3.375 in, height = 3.25 in]{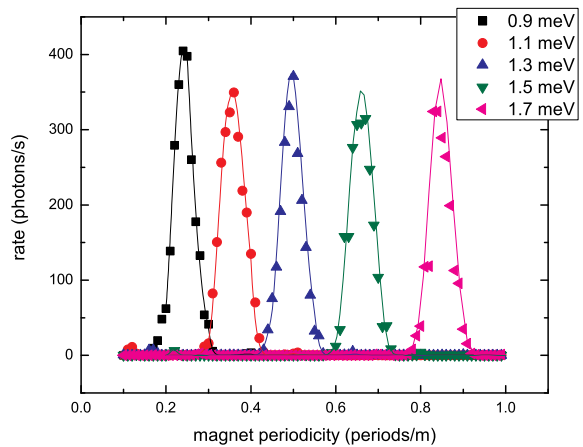}
\caption{Regenerated photon rates from adjustable-period
alternating-field magnet systems, for a coupling scale of
$M_b = 5 \times 10^5$\,GeV.
The magnets are length 10\,m and field amplitude 1.5\,T,
with a 900-nm beam at 10\,kW. The peaks, from left to right,
are for masses of 0.9, 1.1, 1.3, 1.5, and 1.7\,meV.  The points are for a 
segment length of 0.2\,m, the lines for 0.01\,m.}
\label{fig:alternating_rates}
\end{figure}

Figure~\ref{fig:alternating_rates} shows the photon rates of
G-R apparatuses using identical alternating-field magnets.
These calculations used $\delta_k = $\,sgn$[\sin \Delta_m (2k - 1) d/2]$, where $\Delta_m$
is the desired spatial frequency of the field.

Another approach is to use variable-pitch helical magnets as in Figure~\ref{fig:helical};
this design allows better matching to the desired magnet period, and gives similar rates.
However, the azimuthal symmetry does not allow tests of parity, and
spin-zero bosons will generate photons with circular polarization.
For ideal helical magnets
with an incident linearly-polarized beam, the resonant portion of the probability for 
a regenerated photon is

\begin{equation}
P''_{GR} =
\frac{1}{4}\left(\frac{B'_1 \ell_1}{2 M_b}\right)^2\frac{sin^2(y'_1)}{y^{'2}_1}\frac{1}{2}
\left(\frac{B'_2\ell_2}{2 M_b}\right)^2 \frac{sin^2(y'_2)}{y^{'2}_2}
\label{eqn:G-R2}
\end{equation}
where $B'_{1,\,2}$ is the magnitude of the field, and
$\ell_{1,\,2}$ the length, of each magnet,
$y'_{1,\,2} = (\Delta_b - \Delta_{m1,\,2}) \ell_{1,\,2}/2$,
and $\Delta_{m1} = \Delta_{m2} = \Delta_b$ at resonance.
Segmented helical magnets give similar rates
for periods longer than several segments.

Either arrangement would allow
measurement of the boson mass to $\sim\!0.001$\,meV, and the detection of
multiple LNBs within the accessible mass range.
Using light from $10$ to $0.25$\,$\mu$m and periods
from 0.1 to 20\,m, a boson mass range from 0.1 to 10\,meV could be explored,
bracketing the PVLAS range.

If the PVLAS hypothesis is confirmed, a future `LNB factory' could be built
using alternating fields.  To generate and detect LNBs
in the PVLAS region, one could use a magnet length of 100\,m with a field amplitude
of 7\,T.
A 100-m long optical cavity in the production magnet
pumped with 355\,nm laser light could provide an intra-cavity photon flux of 100\,kW
with a Rayleigh range of $>200$\,m~\cite{Hodgson:1997} and work in magnet
apertures of a few cm, giving rates in excess of $10^9$ detected photons/s.

%
\section{\label{sec:studies}Physics studies}

In case of observation of a confirmed signal rate, a first study of
interaction strengths with ordinary matter could be made by adding material
at the wall, and measuring phase shifts or changes in rates.
Given that the energy and wavelength of the LNBs would be in the optical region,
interaction with matter would likely be in the form of coherent forward scattering and
hence could be described by an effective scattering length, providing a connection
with theoretical models.  With a measurement to the accuracy described above (of the
phase shift due to a thick wall), the scattering length per
nucleon or electron could be measured to $\sim\!10^{-10}$\,fm.
Decay or oscillation lengths at the ~1-m scale can be also studied by moving
the regeneration magnet.
The proposed use of an imaging detector that can detect changes in angular
distributions may be relevant to
physics studies.

%
\section{\label{sec:summary}Summary}
Optical wavelengths are well suited to studies of light neutral bosons in the
PVLAS mass and coupling strength region, using the generation-regeneration technique.
High power sources are available, and constant-field and periodic-field magnets
can produce significant fluxes.  Optical detectors with high
efficiency are available, with imaging and timing capabilities.  The expected
coherence of regenerated photons with the incident photons,
assuming very weak coupling to ordinary matter, allow optical
techniques of study to be used.  Some of these techniques will be applied in a forthcoming
experiment by a Hampton University-Jefferson Lab collaboration~\cite{LIPSS:2006a}.

\section{\label{sec:acknowledgements}Acknowledgements}
We acknowledge the help of the  Jefferson Lab FEL staff,
especially Gwyn Williams and George Neil.
Authored in part by The Southeastern Universities Research Association, Inc.
under U.S. DOE Contract No. DE-AC05-84150 and the U.S. Office of Naval Research.
The U.S. Government retains a non-exclusive, paid-up, irrevocable,
world-wide license to publish or reproduce this manuscript for
U.S. Government purposes.
This work was supported in part by U.S. National Science Foundation
awards PHY-0114343 and PHY-0301841.

\bibliography{cosm-06-02}

\end{document}